\documentclass[a4paper]{article}

\usepackage{INTERSPEECH2022}

\usepackage{amsmath,graphicx}
\usepackage{epstopdf}
\usepackage{array}
\usepackage{graphicx}
\usepackage{multirow}
\usepackage{color, soul}
\usepackage{cite}
\usepackage{url}
\usepackage{hyperref}
\usepackage{booktabs}
\usepackage{amsmath}
\usepackage{amssymb}

\title{Accurate Emotion Strength Assessment for Seen and Unseen Speech \\ Based on Data-Driven Deep Learning
}

\name{Rui Liu$^{1,3,4}$, Berrak Sisman$^{4}$, Bj{\"o}rn W.\ Schuller $^{5}$, Guanglai Gao$^{1}$, Haizhou Li$^{2,3}$}
\address{$^1$ Inner Mongolia University, China $^2$ The Chinese University of Hong Kong, Shenzhen, China\\  $^3$  National University of Singapore $^4$ Singapore University of Technology and Design \\ $^5$ Imperial College London, United Kingdom
}

\email{liurui\_imu@163.com, berraksisman@u.nus.edu, csggl@imu.edu.cn,  haizhouli@cuhk.edu.cn}

\begin{document}

\maketitle
\begin{abstract}
Emotion classification of speech and assessment of the emotion strength are required in applications such as emotional text-to-speech and voice conversion. The emotion attribute ranking function based on Support Vector Machine (SVM) was proposed to predict emotion strength for emotional speech corpus. However, the trained ranking function doesn't generalize to new domains, which limits the scope of applications, especially for out-of-domain or unseen speech. In this paper, we propose a data-driven deep learning model, i.e. StrengthNet, to improve the generalization of emotion strength assessment for seen and unseen speech. This is achieved by the fusion of emotional data from various domains. We follow a multi-task learning network architecture that includes an acoustic encoder, a strength predictor, and an auxiliary emotion predictor. Experiments show that the predicted emotion strength of the proposed StrengthNet is highly correlated with ground truth scores for both seen and unseen speech. We release the source codes at: \textcolor{blue}{\href{https://github.com/ttslr/StrengthNet}{https://github.com/ttslr/StrengthNet}}.
\end{abstract}

\section{Introduction}
\label{sec:intro}

Accurate emotion classification of speech and assessment of its strength are essential to profile human behaviors, which has many potential applications, such as human-robot interface, human-machine dialogue, and social media.
In recent years, there is an increasing interest in emotion control in expressive speech synthesis, such as emotional text-to-speech, emotional voice conversion, where accurate control of emotional strength in speech becomes critically important.

The simplest emotion strength control method is to linearly scale the emotion representation vector~\cite{li2021controllable}. The effect of such a linear scale is hardly interpretable.  
To obtain a meaningful strength descriptor, some followed the idea of ``relative attributes''~\cite{zhu2019controlling,lei2021fine,parikh2011relative} and quantify the emotion strength by learning from the $<$neutral, emotional$>$ speech pairs.
Support-Vector-Machine (SVM) based attribute ranking \cite{parikh2011relative} learns the difference between two samples that are significantly different in a particular attribute, that has been widely studied in computer vision \cite{siddiquie2011image,meng2018efficient}.

In speech processing, Zhu et al. \cite{zhu2019controlling} proposed to learn an emotion attribute ranking function $R(\cdot)$ from the $<$neutral, emotional$>$ paired speech features,
then weight the emotional feature with a learnable weighting vector and return a weighted sum as the indicator of the emotion strength of one specific emotional speech.
Lei et al. \cite{lei2021fine} further extended the utterance level emotion attribute ranking function to phoneme level and obtain a fine-grained ranking function.
In this way, we can obtain a meaningful strength score for emotional speech on a specific dataset, which correlates with human perception.

\begin{figure*}[!thp]
\centering
\centerline{ \quad \quad \includegraphics[width=0.9\linewidth]{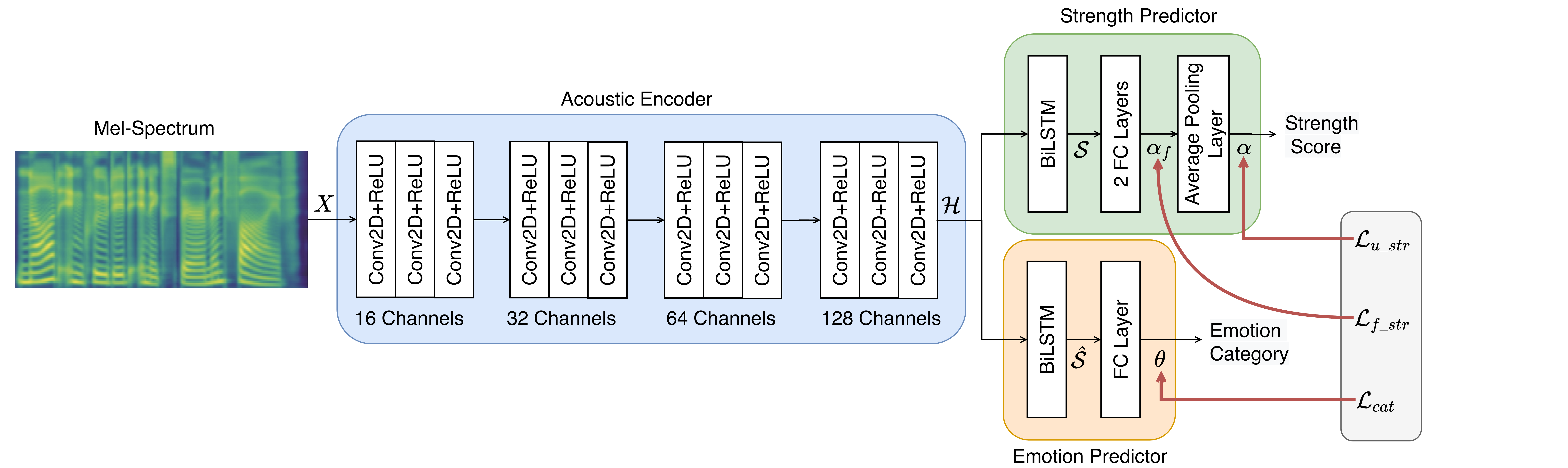}}
\caption{The proposed StrengthNet that consists of Acoustic Encoder, Strength Predictor and Emotion Predictor.}
\label{fig:model}
\end{figure*}

We note that a trained ranking function $R(\cdot)$ on specific data is not easily generalized to new domains. In other words, $R(\cdot)$ is not able to calculate an accurate or appropriate strength score for unseen or out-of-domain speech. To extend to a new data, we need to retrain a new ranking function on the $<$neutral, emotional$>$ paired samples from the new data. Furthermore, the learning of ranking functions for new data requires parallel samples. All these limit the scope of applications. 
Recently, it was shown that deep learning has the ability to learn a mapping function effectively \cite{mosnet,qualitynet}. 
The neural solution learns complex non-linear mapping relationships, and exhibits good generalization ability with the support of a large number of model parameters \cite{zhang2017Understanding}. Most importantly, the data-driven training strategy appears to be more powerful and have the potential to achieve good performance for out-of-domain data \cite{9405474}.

In this paper, we propose a novel neural solution to emotion strength assessment, termed StrengthNet. To improve the model generalization, we employ a data-driven strategy, named \textcolor{black}{``domain fusion'', to mix emotional data from various domains for model training.}
StrengthNet is a multi-task framework that includes a convolutional neural network (CNN) based acoustic encoder, a bidirectional long short-term memory (BiLSTM) based strength predictor and an auxiliary BiLSTM based emotion predictor.
The acoustic encoder extracts the high-level features from the input mel-spectrum. The strength predictor aims to predict the strength score for the input mel-spectrum. The emotion predictor is used to predict the emotion category, which serves as an auxiliary task.

The main contributions of this paper include,
1) We propose a novel data-driven deep learning based speech emotion strength assessment model, i.e., StrengthNet;
2) We show that the predicted emotion strength of seen and unseen speech is highly correlated with the ground truth;
To our best knowledge, this is the first deep learning model for accurate emotion strength assessment for seen and unseen speech.

This paper is organized as follows: 
In Section 2, we formulate the proposed StrengthNet and the domain fusion training strategy. Section 3 report the experimental results. Finally, Section 4 concludes the study.



\section{StrengthNet}
\label{sec:StrengthNet}
We first describe the overall StrengthNet architecture, then explain the domain fusion training details. Lastly, we describe the model inference.

\subsection{Model Overview}
\label{subsec:StrengthNet}

We formulate StrengthNet under the multi-task framework, which consists of an acoustic encoder, a strength predictor and an emotion predictor as in Fig. \ref{fig:model}.

To improve the model generalization ability, we adopt a domain fusion training strategy. During training, the Mel-Spectrum features come from multiple emotional domains. The ground-truth strength scores, which is normalized to (0,1), are derived from a ranking function learned separately.
The details will be described in Section \ref{domainfusion}. We employ a 
 multi-task framework with an auxiliary emotion category prediction task. Next we will describe the model architecture first.

\subsubsection{Acoustic Encoder}
The acoustic encoder takes the acoustic feature sequence, that is mel-spectrum in this work, as input to extract a high-level feature representation.
The acoustic encoder consists of 12 convolution layers.
The strategy of stacking more convolutional layers to expand the receptive field of a CNN has been widely used to model time series data and yield satisfactory performance \cite{mosnet}.
Given an input mel-spectrum sequence $X$, the CNN based acoustic encoder aims to extract a high-level feature $\mathcal{H}$.
The high-level feature $\mathcal{H}$ is then fed to two predictors to predict the emotion strength score and emotion category, respectively.
\subsubsection{Strength Predictor}
The strength predictor then reads the high-level feature representation to predict the emotion strength.
Recent studies have confirmed the effectiveness of combining CNN and BiLSTM for classification \cite{mosnet}, and recognition \cite{9420276} tasks.

The strength predictor consists of a BiLSTM layer, two FC layers, followed by an average pooling layer.
The BiLSTM layer takes the high-level feature $\mathcal{H}$ as input to output the hidden states $\mathcal{S}$ for each time step.
We then use two FC layers to \textcolor{black}{convert} the frame-wise hidden states $\mathcal{S}$ into the frame level scalar $\alpha_{f}$ to indicate the strength score of each frame.
Finally, an average pooling layer is applied to the frame-level scores to obtain the utterance-level strength score $\alpha$.

To supervise the training of the strength predictor, we define a mean absolute error (MAE) loss $\mathcal{L}_{u\_str}$ behind the average pooling layer to force the predicted utterance-level strength score $\alpha$ close to the ground-truth value. 

To improve the convergence of StrengthNet, we further define another MAE loss\cite{mosnet} $\mathcal{L}_{f\_str}$, denoted as frame-wise constraint, behind the last FC layer to minimize the difference between the predicted frame-level strength score $\alpha_{f}$ and the ground-truth strength.

\subsubsection{Emotion Predictor}
The auxiliary emotion predictor is devised to predict the emotion category. Similar to the strength predictor, the emotion predictor also consists of a BiLSTM layer and an additional softmax layer.
The BiLSTM summarizes the temporal information of high-level feature $\mathcal{H}$ into another latent states $\hat{\mathcal{S}}$.
Finally, the softmax layer converts the latent states $\hat{\mathcal{S}}$ to the output probability for all emotion categories.
Accordingly, we can obtain the predicted emotion category $\theta$.
We define a ``categorical\_cross\_entropy'' loss $\mathcal{L}_{cat}$ to restrain the emotion predictor.
We formulate the total objective function $\mathcal{L}_{total}$ for training the StrengthNet as: 
\begin{equation}
    \mathcal{L}_{total} = \mathcal{L}_{f\_str} + \mathcal{L}_{u\_str} + \mathcal{L}_{cat}.
\end{equation}

\subsection{\textcolor{black}{Domain Fusion}}
\label{domainfusion}
\begin{figure}[t]
\centering
\centerline{ \includegraphics[width=0.991\linewidth]{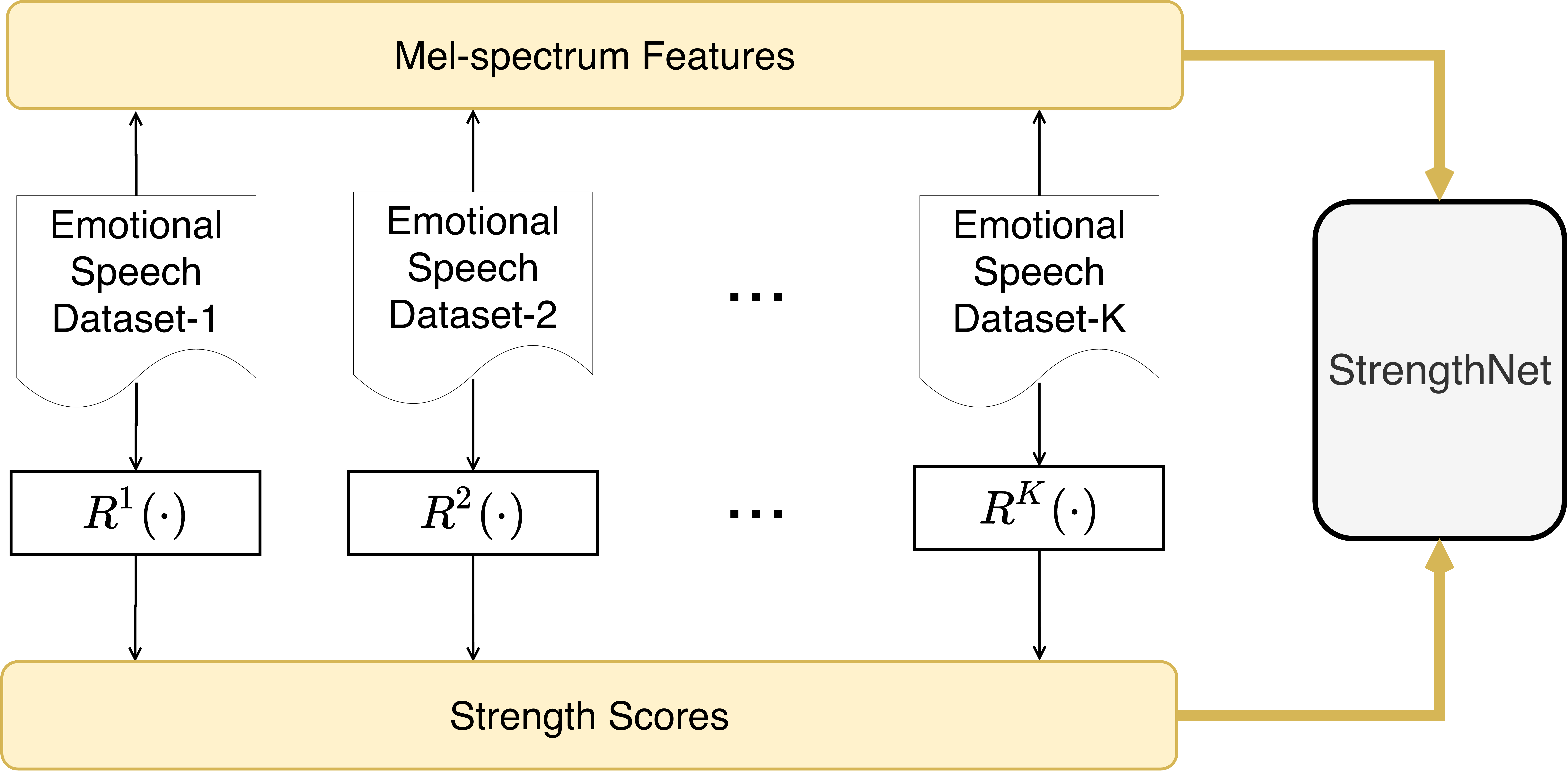}}
\caption{The shown domain fusion strategy is utilized to improve the model generalization.}
\label{fig:dataaug}
\end{figure}

\textcolor{black}{As shown in Fig. \ref{fig:dataaug}, to improve the model generalization, we employ a domain fusion strategy \cite{volpi2018generalizing,shorten2019survey}, which mix multiple emotional speech datasets from various domains to train our StrengthNet. The fused data \textcolor{black}{mixed from various domains} will represent a more comprehensive set, thus minimizing the distance between the training and validation set, as well as any future testing sets. }

Let us denote $\mathcal{D}_{1}$, $\mathcal{D}_{2}$, ..., $\mathcal{D}_{K}$ as the $K$ emotional speech datasets.
We train the emotion attribute ranking function $R(\cdot)$ for all datasets, since they all provide $<$neutral, emotional$>$ pairwise speech samples.

To achieve this, we first build two sets, that are $O$ and $S$, which contain ordered and similar paired samples, respectively.
Specifically, for each dataset $\mathcal{D}_{k}$ ($k \in$ [1,K]), we pick up one sample from neutral speech and another sample from emotional speech to build the ordered set $O$.
We expect that the emotion strength of the emotional sample is higher than that of the neutral sample.
\textcolor{black}{For the similar set $S$, we pick up two samples from `neutral' speech or emotional speech (such as `happy'). We assume that two samples from the same emotion category have more similar emotion strength than two others from different categories.}

We follow \cite{zhu2019controlling} and build a support vector machine (SVM) \cite{joachims2002optimizing} to learn the ranking function $R(\cdot)$ for the emotion strength attribute. 
Finally, we derive the strength scores across all datasets to serve as the ground-truth in the training objective of StrengthNet. Note that the strength score is normalized to (0,1) with 1 having the strongest emotion. We prepare  mel-spectrum features for all datasets as the input of StrengthNet. The StrengthNet can be seen as the aggregation of multiple domain experts.


\subsection{Run-time Inference}
During inference, the StrengthNet takes a mel-spectrum extracted by any emotional speech as the input feature to predict its emotion strength score, as well as its emotion category. Furthermore, our StrengthNet can be used directly to predict emotion strength for a new emotional speech dataset without retraining, that is a clear advantage.


\begin{figure*}[!t]
\centering
\centerline{
\includegraphics[width=0.3\linewidth]{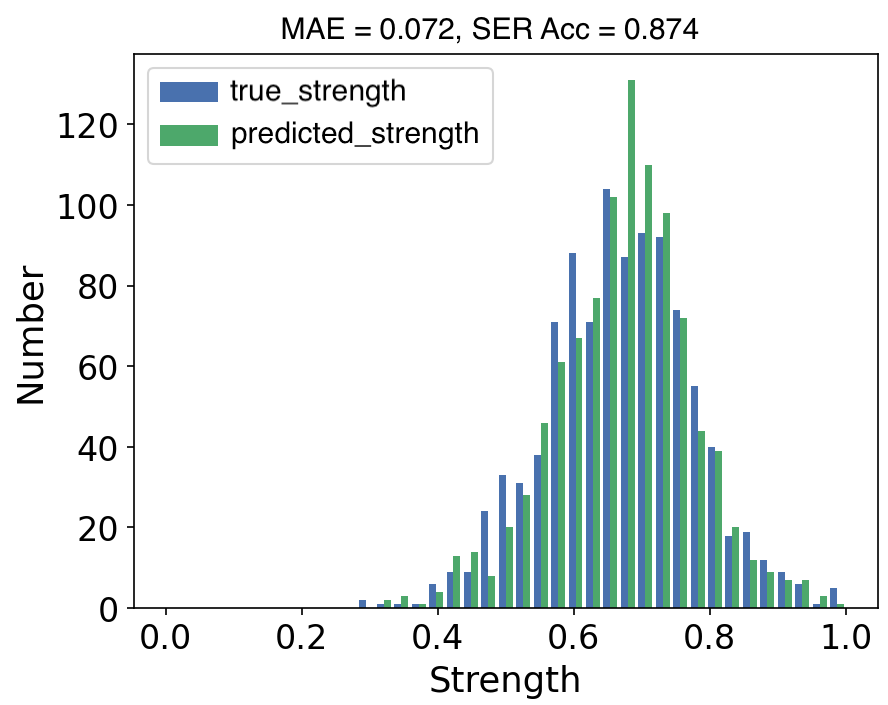}
\includegraphics[width=0.3\linewidth]{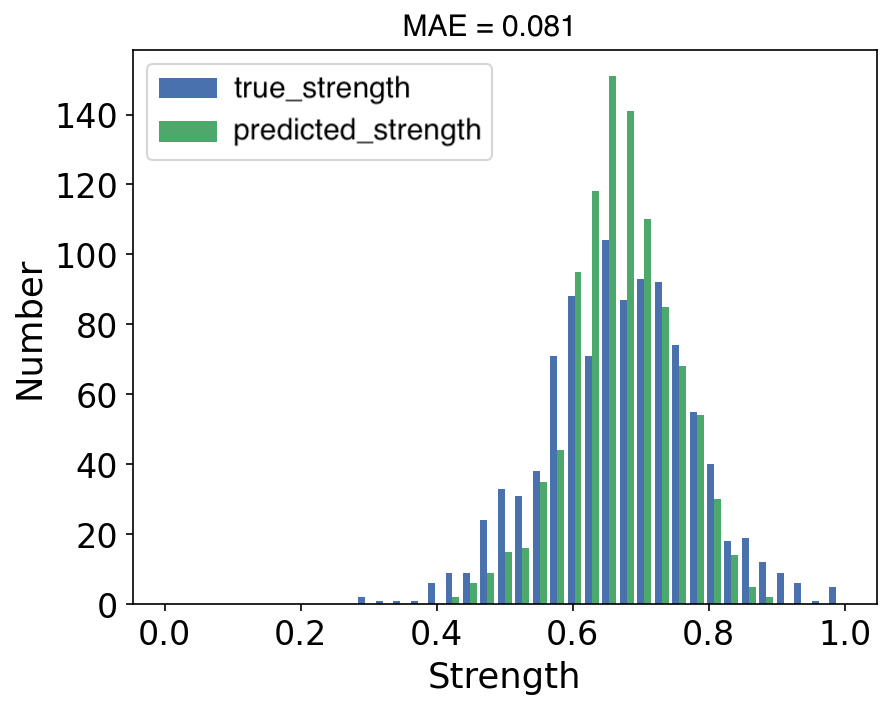}
\includegraphics[width=0.3\linewidth]{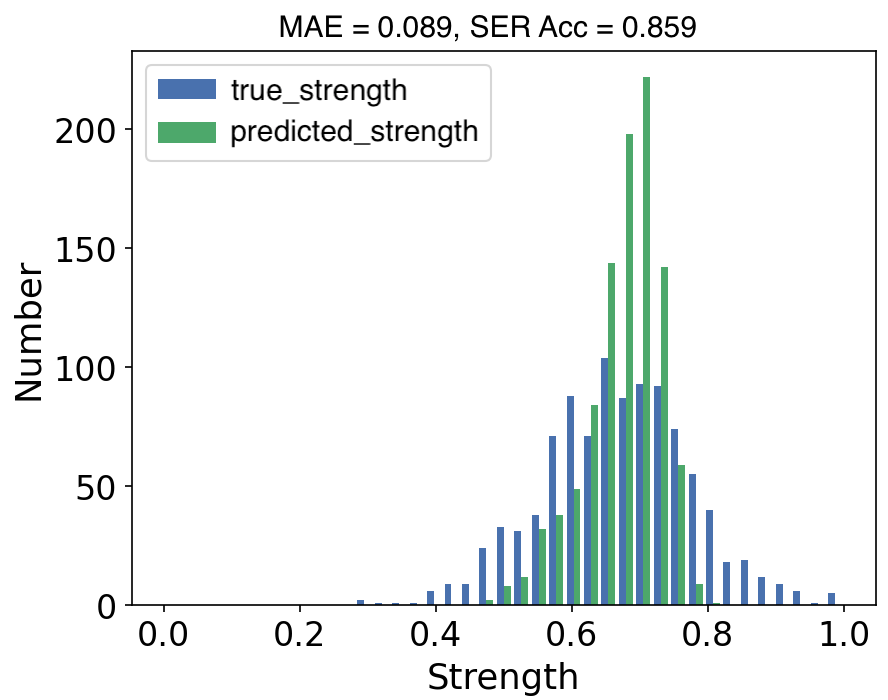}}

\centerline{\quad \quad \quad \quad (a) StrengthNet  \quad \quad \quad \quad \quad \quad \quad \quad   \quad  (b) StrengthNet w/o $\mathcal{L}_{cat}$ \quad \quad \quad \quad \quad  \quad  \quad  \quad  (c) StrengthNet w/o $\mathcal{L}_{f\_str}$ }
\vspace{-1mm}
\caption{Histogram of the utterance level strength predictions for (a) StrengthNet; (b) StrengthNet w/o $\mathcal{L}_{cat}$; (c) StrengthNet w/o $\mathcal{L}_{f\_str}$. The X-axis and Y-axis of subfigures represent the  strength scores and the utterance number, respectively.}
\label{fig:ablation_study}
\vspace{-2mm}
\end{figure*}

\section{Experiments}
\subsection{Datasets}
We use the ESD dataset \cite{zhou2021seen} to validate the performance of StrengthNet in terms of strength prediction.
The English corpus with a total of nearly 13 hours of speech is used in our experiments.
In addition, we use two additional English SER datasets: the Ryerson Audio-Visual Database of Emotional Speech and Song (RAVDESS) \cite{livingstone2018ryerson} and the Surrey Audio-Visual Expressed Emotion (SAVEE) database \cite{jackson2014surrey} to achieve domain fusion and test the model generalization.

The datasets are summarized  as  follows:

\begin{itemize}

\item \textbf{ESD \cite{zhou2021seen}:} ESD is a multilingual emotional speech dataset and has 350 parallel utterances spoken by 10 native English and 10 native Mandarin speakers. We use the English corpora with a total of nearly 13 hours of speech by 5 male and 5 female speakers in five emotions, namely happy, angry, neutral, sad and surprise. All speech samples are sampled at 16 kHz and coded in 16 bits.
\item \textbf{RAVDESS \cite{livingstone2018ryerson}}:
The RAVDESS dataset contains two emotional parts: a speech part and a song part. We mainly focus on its speech part, which has 1440 utterances acted by 24 professional actors (12 female, 12 male). There are 8 classes of emotion contained in the speech part (calm, happy, sad, angry, fear, surprise, disgust, and neutral state). In the process of collecting data, actors are asked to speak two fixed sentences with different classes of emotion. 
\item \textbf{SAVEE \cite{jackson2014surrey}}: The SAVEE database consists of recordings from four male actors in seven different emotions, 480 British English utterances in total with seven emotional states (i.e., surprise, happy, sad, angry, fear, disgust, and neutral). And the sentences are selected from the standard TIMIT corpus and phonetically-balanced for each emotion.

\end{itemize}

\textcolor{black}{
Please note that the emotional speech databases can be divided into two types: acted and improvised \cite{juslin2001impact}. All above three mentioned datasets belongs to acted family which widely used in emotional TTS \cite{cui21c_interspeech}. 
We believe that it is appropriate to employ acted emotional speech dataset for our experiments, due to the improvised speech is hard to induce strong and well-differentiated emotions \cite{banse1996acoustic}.}


\subsection{Experimental Setup}

For all datasets, we take 5 emotion classes, that are happy, sad, angry, surprise, and neutral, to build the $<$neutral, emotional$>$ paired speech to train the ranking function.
During training of StrengthNet, we take 4 emotion classes, that are happy, sad, angry and surprise, into account.

We extract 80-channel mel-spectrum features with a frame size of 50ms and 12.5ms frame shift, that are further normalized to zero-mean and unit-variance, to serve as the model input. 
We employ openSMILE \cite{eyben2013recent} to extract 384-d features for each utterance and \textcolor{black}{follow the relative attributes algorithm \footnote{https://github.com/chaitanya100100/Relative-Attributes-Zero-Shot-Learning} to train the ranking function.} 
The utterance level strength scores for all datasets are obtained using their ranking function and serve as the ground truth score, or training target, for the strength predictor in our StrengthNet.
To calculate the frame level MAE $\mathcal{L}_{f\_str}$, the ground-truth strength score is used for all the frames in the speech utterance.
The emotion predictor of StrengthNet aims to predict 4 emotion categories, including happy, sad, angry and surprise.

The acoustic encoder consists of 4 Conv2D blocks with filters size [16, 32, 64, 128], respectively. Each block includes 3 Conv2D layers with strides shape \{[1,1], [1,1], [1,3]\}, respectively. All layers share the same kernel size [3 $\times$ 3] and ReLU activation function. 
For BiLSTM layers in two predictors, each direction contains 128 cells.

All speech samples are resampled to 16 kHz.
We train the models using the Adam optimizer with a learning rate=0.0001 and $\beta_1$ = 0.9, $\beta_2$ = 0.98. We set the batch size to 64.
The dropout rate is set to 0.3. 
For each dataset, we partition the speech data into training, validation, and test set at a ratio of 8:1:1.
We apply early stopping based on the MAE of the validation set with 30 epochs patience.

\subsection{Experimental Results}

\subsubsection{Architecture Comparison Results}
First, we intend to validate the effectiveness of the multi-task framework and frame constraint in our StrengthNet, in terms of strength prediction performance on the ESD dataset. We develop StrengthNet and two variants for an ablation study, reported as follows: 
1) StrengthNet (\textit{proposed}), which is our proposed model that consists of an acoustic encoder, strength predictor and auxiliary emotion predictor;
2) StrengthNet w/o $\mathcal{L}_{cat}$, which is the proposed model without an auxiliary emotion predictor; and
3) StrengthNet w/o $\mathcal{L}_{f\_str}$, which is the proposed model without a frame strength constraint term $\mathcal{L}_{f\_str}$ within the strength predictor.

\textcolor{black}{We report the performance in terms of MAE values. Note that the MAE can range from 0 to $\infty$. It is negatively-oriented scores and lower values are better.}
Fig. \ref{fig:ablation_study} shows the overall performance of these systems at the utterance level. We report the MAE value of emotion strength prediction and the accuracy of emotion category prediction on test data.
We observe in Fig. \ref{fig:ablation_study} that our proposed StrengthNet outperforms Strength w/o $\mathcal{L}_{cat}$ and Strength w/o $\mathcal{L}_{f\_str}$ and achieves the best performance, that is attributed to the multi-task and frame constraint strategy.  Specifically, we find that StrengthNet achieves the lowest MAE score of 0.072 and highest emotion recognition accuracy (denoted as \textit{SER Acc} \textcolor{black}{as shown in the axis titles of Fig. \ref{fig:ablation_study}}) of 0.874.
To sum up, the above results indicate that the combination of multi-task learning and frame constraint can effectively learn emotion strength cues from the input mel-spectrum to perform emotion strength prediction well, along with category prediction.

\begin{table}
\centering
\caption{MAE results for the RAVDESS and SAVEE datasets in the comparison.}
\vspace{-3mm}
\begin{tabular}{lcc}
\toprule
\multirow{2}{*}{\textbf{Method}}& \multicolumn{2}{c}{\textbf{MAE}}\\  
  &  RAVDESS & SAVEE  \\
 \bottomrule
$R_{{\rm RAVDESS}}(\cdot)$ & NA & 0.304 \\ \cline{1-3} 
$R_{{\rm SAVEE}}(\cdot)$ & 0.283 & NA \\ \cline{1-3} 
$R_{{\rm ESD}}(\cdot)$ & 0.266 & 0.272 \\ \cline{1-3} 
$\rm StrengthNet_{{\rm ESD}}$ & 0.238 & 0.243 \\ \cline{1-3} 
$\rm StrengthNet_{{\rm ESD+RAVDESS}}$ & NA & \textbf{0.173} \\ \cline{1-3} 
$\rm StrengthNet_{{\rm ESD+SAVEE}}$ & \textbf{0.102} & NA \\ 
\bottomrule
\end{tabular}
\label{tab:dataaug}
 \vspace{-4mm}
\end{table}

\vspace{-2mm}
\subsubsection{Domain Fusion Results}
\vspace{-1mm}

We further verify the effectiveness of the proposed domain fusion for model generalization in terms of emotion strength prediction.

\begin{sloppypar}
We train StrengthNet on three dataset settings, that are ESD, ESD+RAVDESS, ESD+SAVEE. 
$\rm  StrengthNet_{ESD}$, $\rm StrengthNet_{ESD+RAVDESS}$ and $\rm StrengthNet_{ESD+SAVEE}$ are used to represent three trained models, respectively.
$R_{{\rm RAVDESS}}(\cdot)$, $R_{{\rm SAVEE}}(\cdot)$ and $R_{{\rm ESD}}(\cdot)$ refer to the three trained ranking functions for three datasets, respectively. 
We adopt all trained ranking functions and models to predict the emotion strength score on unseen test data from RAVDESS or SAVEE, and report the MAE results in Table \ref{tab:dataaug}.
\end{sloppypar}

As shown in Table \ref{tab:dataaug}, we observe that our StrengthNet performs a lower MAE than ranking functions on both, the RAVDESS and SVAEE datasets. More importantly, in the case of $\rm StrengthNet_{ESD+RAVDESS}$ and $\rm  StrengthNet_{ESD+SAVEE}$, the MAE achieves the lowest values of 0.173 and 0.102 on SVAEE and RAVDESS respectively.
As can be seen from the results, our proposed strength method can reduce the overall MAE on unseen data with the domain fusion strategy, which performs better model generalization than the attribute ranking function.

\vspace{-1mm}
\subsubsection{\textcolor{black}{Human Perception Evaluation}}
\vspace{-1mm}

\textcolor{black}{
We further evaluate how close the results of StrengthNet for unseen speech can get to the emotional strength as perceived by humans.}

\textcolor{black}{
Since the RAVDESS dataset includes the manual emotion intensity label, which each emotion is produced at two levels of emotion intensity: normal and strong. Therefore, we adopt $R_{{\rm SAVEE}}(\cdot)$, $R_{{\rm ESD}}(\cdot)$ and $\rm  StrengthNet_{ESD+SAVEE}$ to predict the emotion intensity for the test set of RAVDESS.
We consider the predicted intensity scalars from 0 to 0.5 as `normal' and 0.5 to 1.0 as `strong' in two categories. We select 20 utterances from the test set.}

 \begin{figure}[t]
\centering
 \centerline{
    \includegraphics[width=0.35\linewidth]{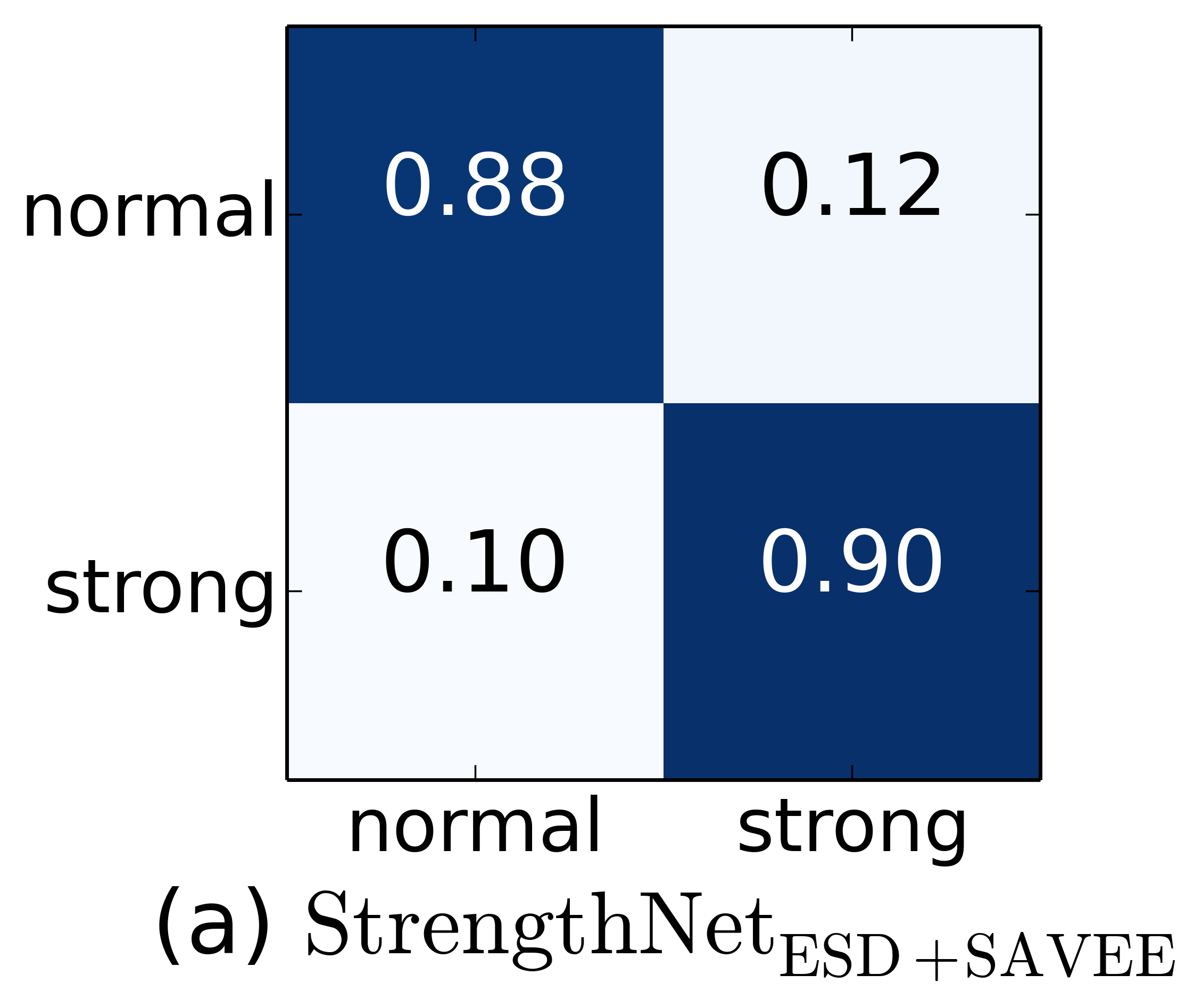}
    \includegraphics[width=0.31\linewidth]{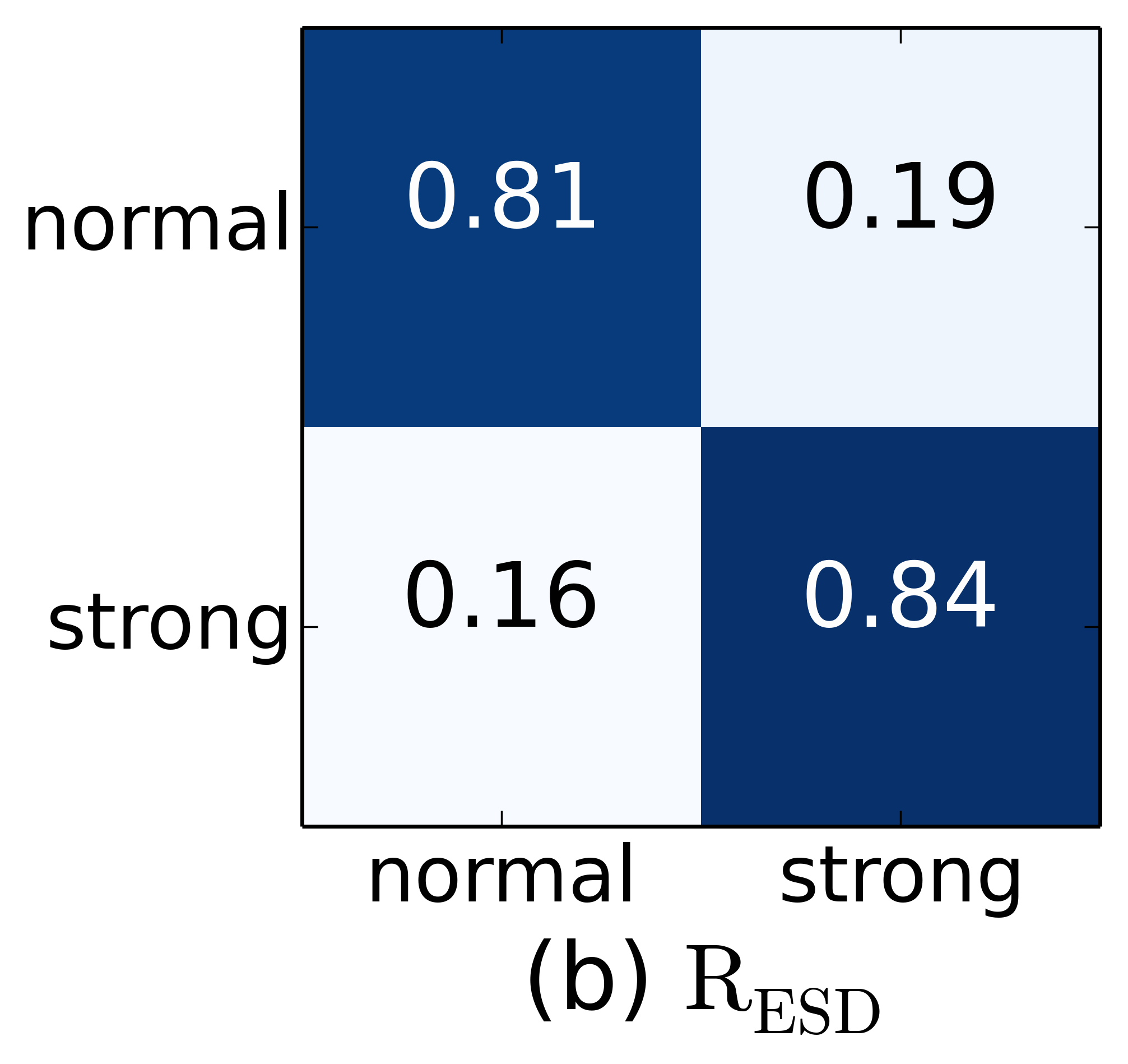}
     \includegraphics[width=0.31\linewidth]{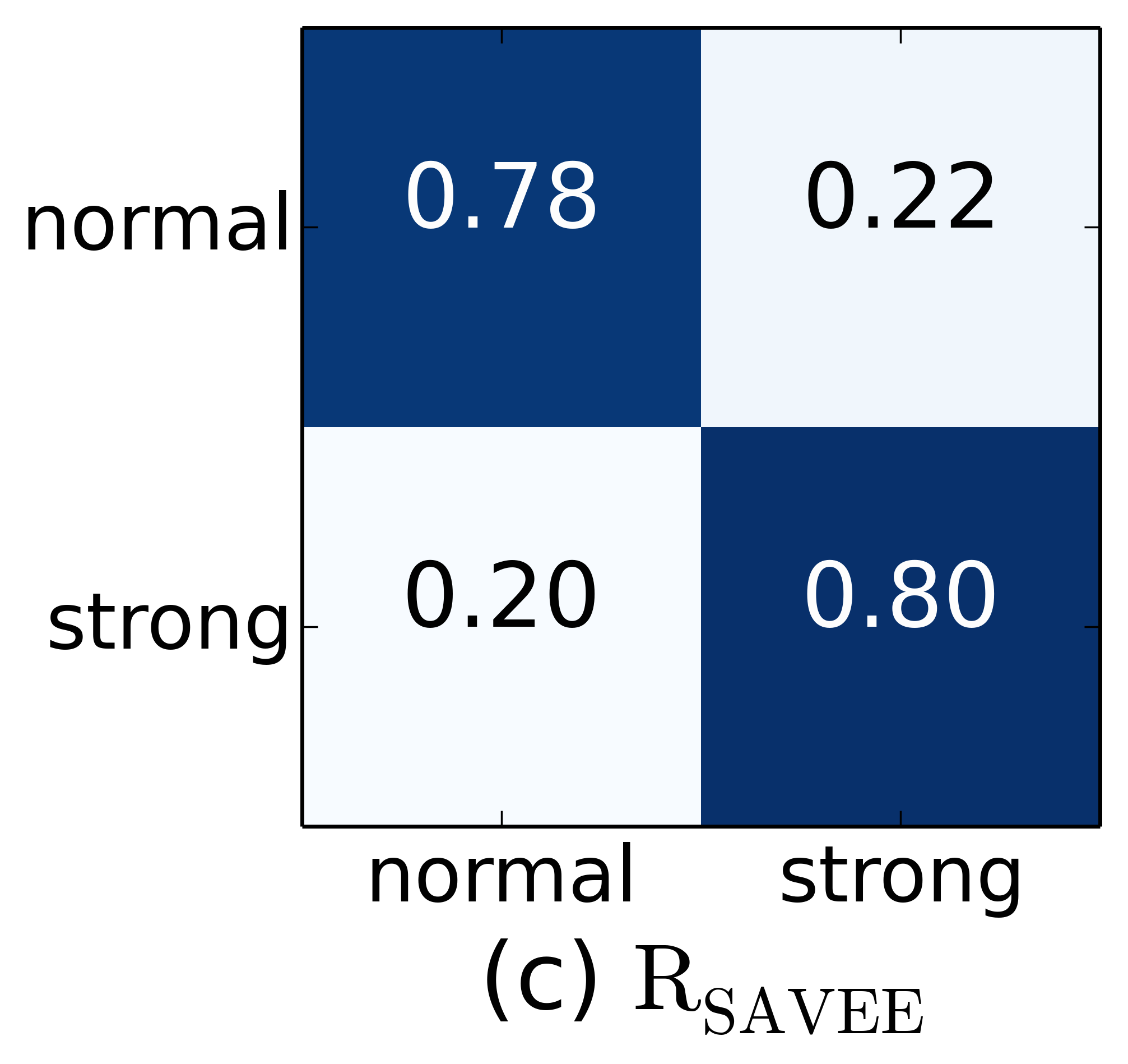}
    }
\vspace{-3mm}\caption{Confusion matrices between predicted and perceived emotion strength categories for unseen speech. 
The X-axis and Y-axis of the figures represent the predicted and perceived category, namely ``normal'' and ``strong''.}
\label{fig:matirx1}
\vspace{-5mm}
\end{figure}

\textcolor{black}{
Fig. \ref{fig:matirx1} presents the intensity confusion matrices.
It is observed that the StrengthNet shows a higher correlation between the predicted and perceived emotion intensity categories, with a correlation of over 85\%, that is considered a competitive result against the trained ranking function. The experiments confirm the superiority of the proposed domain fusion based StrengthNet for unseen speech in terms of human perception.}

\section{Conclusion}
This paper presents a data-driven deep learning-based speech emotion strength assessment model for the emotional speech synthesis task, referred to as StrengthNet.
Experimental results demonstrate that our StrengthNet can achieve accurate emotion strength prediction for both seen and unseen speech with the help of domain fusion strategy.
As per our knowledge, the proposed StrengthNet is the first end-to-end speech emotion strength assessment model.
In future work, we intend to integrate our StrengthNet as a front-end or back-end for emotional speech synthesis models to enhance the emotion expressiveness of output emotional speech.



\bibliographystyle{IEEEtran}

\bibliography{mybib}


\end{document}